\documentclass[aps,preprint,a4paper, showkeys]{revtex4-1}
\usepackage{graphicx}
\usepackage{amsmath}
\usepackage{times}
\usepackage{color}
\usepackage[latin1]{inputenc}
\def\nn{\nonumber}
\def\beq{\begin{equation}}
\def\eeq{\end{equation}}
\def\beqna{\begin{eqnarray}}
\def\eeqna{\end{eqnarray}}
\def\bea{\begin{array}}
\def\ea{\end{array}}

\def\MU{{\mathcal U}}
\def\MV{{\mathcal V}}

\begin{document}
\title{Signal to noise ratio in parametrically-driven oscillators}
\author{Adriano A. Batista and Raoni S. N. Moreira}
\affiliation{
Departamento de Física\\
Universidade Federal de Campina Grande\\
Campina Grande-PB\\
CEP: 58109-970\\
Brazil}
\date{\today}
\begin{abstract}
Here we report a theoretical model based on Green's functions and
averaging techniques that gives analytical estimates to the signal to noise
ratio (SNR) near the first parametric instability zone in
parametrically-driven oscillators in the presence of added ac drive and
added thermal noise. 
The signal term is given by the response of the parametrically-driven
oscillator to the added ac drive, while
the noise term has two different measures: one is dc and the other is
ac. The dc measure of noise is given by a time-average of the statistically-averaged
fluctuations of the position of the parametric oscillator due to thermal noise.
The ac measure of noise is given by the amplitude of the
statistically-averaged fluctuations at the frequency of the parametric pump. We
observe a strong dependence of the SNR on the phase between the external drive
and the parametric pump, for some range of the phase there is a high SNR, while
for other values of phase the SNR remains flat or decreases with increasing pump
amplitude. Very good agreement between analytical estimates and numerical
results is achieved.
\end{abstract}
\maketitle
\section{Introduction}
Parametrically-driven systems and parametric resonance occur in many different
physical systems, ranging from Faraday waves \cite{faraday1831}, inverted
pendulum stabilization, stability of boats, balloons, and parachutes \cite{ruby1996}. More recent applications in micro and
nano systems include quadrupole ion guides and ion traps \cite{paul90},opto-mechanical
cavities \cite{dorsel83}, magnetic resonance force microscopy
\cite{dougherty1996detection}, tapping-mode force microscopy
\cite{moreno2006parametric}, axially-loaded microelectromechanical devices
(MEMS) \cite{requa06electromechanical}, and torsional MEMs
\cite{turner98nature}, just to mention a few
relevant applications.

Parametric pumping has had many applications in the field of MEMs, which have
been used primarily for measuring small forces and as ultrasensitive mass
detectors since the mid 80's \cite{binnig86}. An enhancement to the detection
techniques in MEMs was developed in the early 90's that uses mechanical
parametric amplification (before transduction) to improve the sensitivity of measurements.  This
amplification method works by driving the parametrically-driven resonator on
the verge of parametric unstable zones.  Rugar and Grütter \cite{rugar91} have
shown ways, using this method, to obtain linear parametric gain.  
Furthermore, while they were
looking for means of reducing noise and increasing precision in a detector
for gravitational waves, they experimentally found classical thermomechanical 
quadrature squeezing, a phenomenon which is reminiscent of quantum squeezed states.
Further experimental studies of parametric amplification appeared in~\cite{almog2007noise}, where a linear 
response around a limit cycle due to 
noise yields noise squeezing in a driven Duffing oscillator.
Implementations of parametric oscillators in electronic circuits can be found in
Refs.~\cite{falk1979,berthet2002}. Parametric amplification started being
studied in electronic systems in the late 50's and early 60's by P. K. Tien
\cite{tien1958parametric}, R. Landauer \cite{landauer1960parametric}, and
Louisell \cite{louis60}. It has been used for its desirable characteristics of
high gain and low noise. Recent applications of parametric amplification in
electronics can be found in Ref.~\cite{lee2011low}.

The limits of parametric amplification  due to thermomechanical noise 
on parametric sensing of small masses in nanomechanical oscillators 
have been studied in
\cite{cleland_njp05}. Although this work is quite broad the author 
does not provide estimates for the SNR as one approaches the first instability
zone (e.g. by increasing the pump amplitude).
Here, we study the effect of adding thermal noise to a parametrically-driven
oscillator with the objective of studying the effectiveness of parametric
amplification in the presence of noise. This one-degree of freedom model may be
applied for instance to the fundamental mode of a doubly-clamped beam resonator
that is axially loaded, in which case the one degree of freedom represents
the amount of deflexion of the middle of the beam from the equilibrium position.
The present model can also be applied to the linear response of driven
nonlinear oscillators to noise (such as transversally-loaded beam resonators),
see for example Ref.~\cite{almog2007noise}. One of us (A.A.B.)
recently obtained analytical quantitative estimates of the amount of quadrature
noise squeezing, heating or cooling in a parametrically-driven oscillator
\cite{batista2011cooling}.
We now use the Green's function approach, previously developed to solve the
Langevin equation, aligned with averaging techniques, to obtain analytical
estimates of the signal-to-noise ratio (SNR) in the parametrically-driven
oscillator in the presence of both added noise and external sinusoidal drive.
Here we show that for some values of phase (between the pumping and external
drives), the signal grows faster than the fluctuations due to added noise, while
for some other values of phase, the SNR is flat or decreases as the pumping
amplitude grows, when one gets close to first instability zone of parametric
resonance.
\section{Theoretical model}
\begin{figure}[!h]
    \includegraphics[scale=0.8]{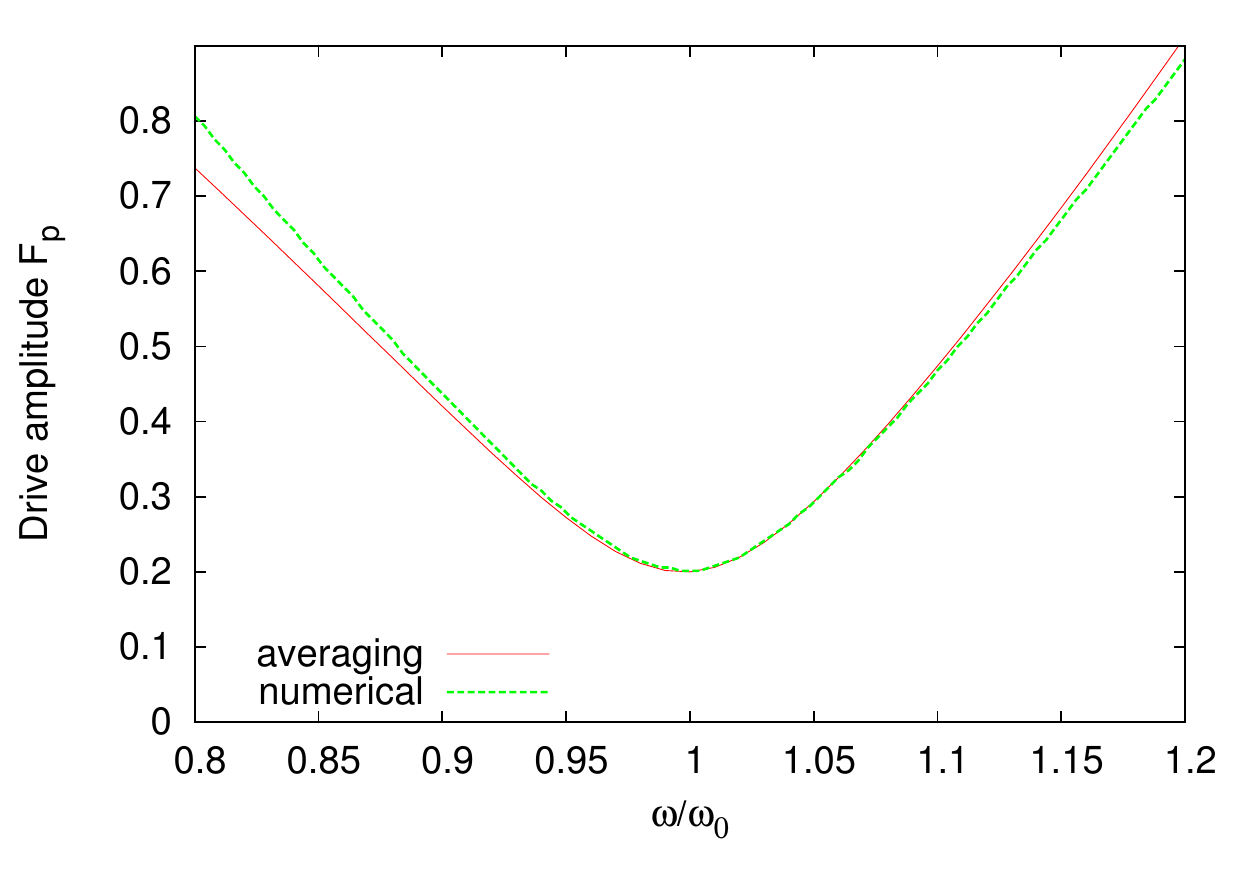} 
    \caption{ (color online) Comparison
    between numerical and averaging method predictions for the boundary of the
    first instability zone of the damped parametrically-driven oscillator of
    Eq.~(\ref{parOsc}). In the region above the green line lies the unstable
    zone obtained by numerical computation, while the region above the red line
    is the analytical prediction for the unstable zone. The numerical results
    are obtained by numerically calculating the corresponding Floquet
    multipliers, when at least one of them has modulus equal to 1. The averaging
    predictions are given by Eq.~(\ref{eq_zones}). The fixed parameters of the
    equations of motion are $\gamma=0.1,\omega_0=1.0$.  These parameters were
    also used to obtain the results portrayed in the remaining figures.  }
    \label{zones}
\end{figure}

\noindent
The equation for the parametrically-driven  oscillator (in dimensionless format) is given by
the damped Matthieu's equation
\begin{equation}
    \ddot x+\omega^2_0x=-\gamma\dot x+F_p\cos(2\omega t)\;x,
    \label{parOsc}
\end{equation}
in which $\gamma$ and $F_p\sim O(\varepsilon)$, where $\varepsilon<<1$.
Since we want to apply the averaging method (AM) \cite{verh96, Guck83} to
situations in which we have detuning, it is convenient to rewrite
Eq.~(\ref{parOsc}) in a more appropriate form with the
notation $\Omega=\omega^2_0-\omega^2$, where we also have $\Omega\sim O(\varepsilon)$.
With this substitution we obtain
$\ddot x+\omega^2x=-\Omega x-\gamma\dot x+F_p\cos(2\omega t)\;x$.
We then rewrite this equation in the form
$\dot x = y$, $\dot y = -\omega^2x+f(x,y,t)$,
where $ f(x,y,t)= -\Omega x+F_p\cos(2\omega t)\;x-\gamma y$.
We now set the above equation in slowly-varying form with the transformation to a slowly-varying frame
\beq
\left(\bea{c}
x \\ y
\ea
\right)
=
\left(
\begin{array}{cc}
    \cos\omega t & -\sin \omega t\\
    -\omega\sin\omega t & -\omega\cos\omega t
\end{array}
\right)
\left(\bea{c}
\MU \\ \MV
\ea
\right)
\label{slow_trans}
\eeq
   and obtain 
\beqna
\left(\bea{c}
\dot \MU \\ \dot \MV
\ea
\right)
&=&
\left(
\begin{array}{cc}
    \cos \omega t & -\frac{1}{\omega}\sin\omega t\\
    -\sin\omega t & -\frac{1}{\omega}\cos\omega t\nn
\end{array}
\right)
\left(\bea{c}
0\\  f(x,y,t)
\ea
\right)\nn\\
&=&
-\frac{1}{\omega}\left(\bea{c} \sin\omega t f(x,y,t)\\\cos\omega t f(x,y,t)\ea\right).\nn
\eeqna
After an application of the AM (in which, basically,
we filter out oscillating terms at and near $2\omega$ in the above equation), we obtain
\begin{eqnarray}
    \dot {\tilde u} &=& \frac{-1}{2\omega}\left[\gamma\omega \tilde u
    +\left(\Omega +\frac{F_p}{2}\right)\tilde v\right],\nn\\
    \dot {\tilde v} &=& \frac{-1}{2\omega}\left[\left(-\Omega +\frac{F_p}{2}\right)\tilde u+\gamma\omega \tilde v\right],\nn
\end{eqnarray}
where the functions $\MU(t)$ and $\MV(t)$ were replaced by their slowly-varying
averages $\tilde u(t)$ and $\tilde v(t)$, respectively. 
The averaging theorem \cite{Guck83} tells that these two sets of functions will
be close to each other to order $O(\epsilon)$ during a time scale of $O(1/\epsilon)$ if
they have initial conditions within an initial distance of $O(\epsilon)$.
So by studying the simpler averaged system, one may obtain very accurate
information about the corresponding more complex non-autonomous original system.
With the transformations $\tilde u(t)= e^{-\gamma t/2} u(t)$ and $\tilde v(t)=
e^{-\gamma t/2} v(t)$, we obtain
\begin{eqnarray}
    \dot u &=& \frac{-1}{2\omega}\left(\Omega +\frac{F_p}{2}\right) v,\nn\\
    \dot v &=& \frac{-1}{2\omega}\left(-\Omega +\frac{F_p}{2}\right) u,
    \label{noise_av}
\end{eqnarray}
Upon integration of
Eq.~(\ref{noise_av}), one finds the solution
\beqna
\tilde u(t)&=&e^{-\gamma t/2}\left[u_0 \cosh (\kappa
t)+\frac{\beta -\delta}{\kappa}v_0\sinh(\kappa t)\right]\nn,\\
\tilde v(t)&=&e^{-\gamma t/2}\left[v_0 \cosh (\kappa
t)+\frac{\beta+\delta}{\kappa}u_0 \sinh(\kappa t)\right],
\label{noise_sol}
\eeqna
where $\kappa=\sqrt{\beta^2-\delta^2}$, $\beta=-F_p/4\omega$, and 
$\delta=\Omega/2\omega$. Hence, we find that the first parametric resonance, i.e. the boundary between
the stable and unstable responses, is given by 
\beq
(\gamma\omega)^2=(F_p/2)^2-\Omega^2.
\label{eq_zones}
\eeq
This result is valid for $\omega\approx \omega_0$ even in the presence of
added noise. In Fig.~(\ref{zones}) we find very good agreement between results
obtained from numerical integration of Eq.~(\ref{parOsc}) and the boundary given by the
averaging technique.

We now will investigate the effect of added thermal noise on the parametric amplification mechanism
\cite{rugar91, difilippo92prl, natarajan95prl, almog2007noise}.
We start by adding noise to Eq.(\ref{parOsc}) and obtain
\begin{equation}
    \ddot x=-\omega^2_0 x-\gamma\dot x+F_p\cos(2\omega t)\;x+R(t),
    \label{DuffingNoise}
\end{equation}
where $R(t)$ is a random function that 
satisfies the statistical averages
$\langle R(t)\rangle=0$
and $\langle R(t)R(t')\rangle= 2T\gamma\delta(t-t')$, according to the
fluctuation-dissipation theorem \cite{kubo66}. $T$ is the temperature of the
heat bath in which the oscillator (or resonator) is embedded. Once we integrate
these equations of motion we can show how classical mechanical noise squeezing,
heating, and cooling occur. We now summarize the method developed in
\cite{batista2011cooling} to analytically study the parametrically-driven
oscillator with added noise, as given by Eq.~(\ref{DuffingNoise}).

\subsection{Green's function method}

The equation for the Green's function of the parametrically-driven oscillator
is given by
\begin{equation}
    \left[\frac{\partial^2}{\partial t^2} +\omega_0^2+\gamma \frac{\partial}{\partial t}-F_p\cos(2\omega t)\right]\;G(t,t')=\delta(t-t').
    \label{green_eq}
\end{equation}
Since we are interested in the stable zones of the parametric oscillator,
for $t<t'$ $G(t,t')=0$ and by integrating the above equation near $t=t'$, 
we obtain the initial conditions when $t=t'+0^+$, $G(t,t')=0$ and $\frac{\partial}{\partial t}G(t,t')=1.0$.
Using the Green's function we obtain the solution $x(t)$ of 
Eq.~(\ref{DuffingNoise}) in the presence of noise $R(t)$
\begin{align}
x(t)&= x_h(t)+\int_{-\infty}^{\infty}dt'\;G(t,t')R(t'),\\
\dot x(t)&=v(t)= v_h(t)+\int_{-\infty}^{\infty}dt'\;\frac{\partial}{\partial t}G(t,t')R(t'),
\end{align}
where $x_h(t)$ is the homogeneous solution, which in the stable zone decays
exponentially with time; since we assume the pump has been turned on for a long
time, $x_h(t)=0$.
By statistically averaging the  fluctuations as a function of time
we obtain
\beqna
\langle x^2(t)\rangle &=& \iint_{-\infty}^{\infty}dt'\;dt''G(t,t')G(t,t'')\langle R(t')R(t'')\rangle=2T\gamma\int_{0}^{\infty}d\tau\;G(t,t-\tau)^2,
\label{x2_avg}\\
\langle v^2(t)\rangle
&=&2T\gamma\int_{-\infty}^{t}dt'\;\left[\frac{\partial}{\partial
t}G(t,t')\right]^2\nn\\
 &=& \omega^2_0\langle x^2(t)\rangle-2F_p\gamma
 T\int^{\infty}_{0}d\tau\cos(2\omega (t-\tau))\;G(t-\tau, t)^2.
\label{v2_avg}
\eeqna
where $\tau=t-t'$.

Although equation (\ref{DuffingNoise}) may be solved exactly by using Floquet 
theory and Green's functions methods \cite{wies85}, one obtains very complex 
 solutions. 
Instead, we find fairly simple analytical approximations to the
Green's functions and, subsequently, to the statistical averages of fluctuations using the
averaging method.
We then use the solution of the system of coupled differential equations
(\ref{noise_sol}), where the initial conditions at $t=t'$ are given by
$u(t')=-\sin(\omega t')/\omega$ and $v(t')=-\cos(\omega t')/\omega$ and obtain
the approximate Green's function is
\beqna
&&G(t,t')\approx -\frac{e^{-\gamma (t-t')/2}}{\omega}\left[\cos(\omega t)\left(\cosh(\kappa\,(t-t'))\sin(\omega t')+\frac{\beta-\delta}{\kappa}\sinh(\kappa\,(t-t'))\cos(\omega t')\right)\right.\nn\\
&&\left.-\sin(\omega t)\left(\frac{\beta+\delta}{\kappa}\sinh(\kappa\,(t-t'))\sin(\omega t')+\cosh(\kappa\,(t-t')) \cos(\omega t')\right)\right],
\label{green_avg}
\eeqna
for $t>t'$ and $G(t,t')=0$ for $t<t'$.
In the stable zone of the parametrically-driven oscillator, when
$|\beta|>|\delta|$ , we can rewrite the Green's function replacing the initial conditions and
using simplifying trigonometrical identities. The change of variables $t'=t-\tau$ leads to
\beqna
&&G(t,t-\tau)\approx \frac{e^{-\gamma \tau/2}}{\omega}
\left\{
\cosh(\kappa\,\tau)\sin(\omega \tau)
+\frac{\delta}{\kappa}\sinh(\kappa\,\tau)\cos(\omega \tau)\right.\nn\\
&&\left.-\frac{\beta}{\kappa}\sinh(\kappa\,\tau)\left[\cos(\omega
\tau)\cos(2\omega t)+\sin(\omega \tau)\sin(2\omega t)\right] \right\}.
\label{green_avg2}
\eeqna
We notice that by varying the pump amplitude $F_p$ and the detuning $\Omega$,
we can create a continuous family of classical thermo-mechanical squeezed states,
generalizing the experimental results of Rugar and Gr\"utter \cite{rugar91}.
An estimate of the time average of the thermal fluctuations, when $|\beta|>|\delta|$, is given by
\beqna
\overline{\langle x^2(t)\rangle}&=&\frac{2T\gamma}{\omega^2}\int_0^\infty e^{-\gamma\tau}\left\{\left[\cosh(\kappa\tau)\sin(\omega\tau)+\frac{\delta}{\kappa}\sinh(\kappa\tau)\cos( \omega\tau)\right]^2+\frac{\beta^2}{2\kappa^2}\sinh^2(\kappa\tau)\right\}\,d\tau\nn\\
&=&\frac{2T\gamma}{\omega^2}
\left[I_1+I_2+\delta^2I_3+ \delta I_4\right],
\label{x2_avg_dc2}
\eeqna
where the integrals are given by
\beqna
I_1&=&\frac{\beta^2}{2\kappa^2}\int_0^\infty e^{-\gamma\tau}\sinh^2(\kappa\tau) d\tau=
\frac{\beta^2}{\gamma(\gamma^2-4\kappa^2)},\nn\\
I_2&=&\int_0^\infty e^{-\gamma\tau}\cosh^2(\kappa\tau) \sin^2(\omega\tau) d\tau\nn\\
&=&\frac{1}{2}\left\{\frac{1}{2\gamma}-\frac{\gamma}{2(\gamma^2+4\omega^2)}
+\frac{\gamma}{2(\gamma^2-4\kappa^2)}-\frac{1}{4}\mbox{Re}\left[\frac{1}{\gamma-2\kappa-2i\omega}+\frac{1}{\gamma+2\kappa-2i\omega}\right]\right\},\nn\\
I_3&=&\frac{1}{\kappa^2}\int_0^\infty e^{-\gamma\tau}\sinh^2(\kappa\tau) \cos^2(\omega\tau) d\tau\nn\\
&=&\frac{1}{\kappa^2}\left\{\frac{\kappa^2}{\gamma(\gamma^2-4\kappa^2)}+\frac{1}{8}\mbox{Re}\left[\frac{1}{\gamma-2\kappa-2i\omega}+\frac{1}{\gamma+2\kappa-2i\omega}\right]-\frac{\gamma}{4(\gamma^2+4\omega^2)}\right\},\nn\\
\label{I_3}
I_4&=&\frac{1}{2\kappa}\int_0^\infty e^{-\gamma\tau}\sinh(2\kappa\tau)\sin(2\omega\tau) d\tau=\frac{1}{4}\mbox{Im}\left[ \frac{1}{\kappa(\gamma-2\kappa-2i\omega)}-\frac{1}{\kappa(\gamma+2\kappa-2i\omega)}\right].\nn
\eeqna
A time-averaged estimate of the statistically averaged thermal fluctuations in
velocity, when $|\beta|>|\delta|$, is given by
\beqna
\overline{ \langle v^2(t)\rangle} &=& \omega^2_0\overline{ \langle x^2(t)\rangle}
+\frac{4F_pT\beta}{\kappa}\int^{\infty}_{0}d\tau\,e^{-\gamma\tau}\sinh(\kappa\,\tau)\left[\cosh(\kappa\,\tau)\sin(\omega \tau)
+\frac{\delta}{\kappa}\sinh(\kappa\,\tau)\cos(\omega \tau)\right]\times\nn\\
&&\overline{\cos(2\omega (t-\tau))\left[\cos(2\omega (t-\tau))\cos(\omega \tau)+\sin(2\omega (t-\tau))\sin(\omega \tau)\right]}\nn\\
&=& \omega^2_0\overline{ \langle x^2(t)\rangle}-8\omega T\beta^2(\delta
I_3+I_4).
\label{v2_avg_dc}
\eeqna

An estimate of the statistically averaged thermal fluctuations, when $|\beta|>|\delta|$, is given by

\beqna
\langle x^2(t)\rangle&\approx&\overline{\langle x^2(t)\rangle}+A_{2\omega}\cos(2\omega t)+B_{2\omega}\sin(2\omega t)+A_{4\omega}\cos(4\omega t)
+B_{4\omega}\sin(4\omega t),
\label{x2_avg2}
\eeqna
where
\beqna
A_{2\omega}&=& -\frac{4\beta T\gamma}{\omega^2}(K_1+K_2),\nn\\
B_{2\omega}&=& -\frac{4\beta T\gamma}{\omega^2}(K_3+K_4),\nn
\eeqna
with
\beqna
K_1&=&
\frac{1}{8}\mbox{Im}\left[ \frac{1}{\kappa(\gamma-2\kappa-2i\omega)}-\frac{1}{\kappa(\gamma+2\kappa-2i\omega)}\right],\nn\\
K_2&=&\frac{\delta}{\kappa^2}\left\{\frac{\kappa^2}{\gamma(\gamma^2-4\kappa^2)}-\frac{\gamma}{4(\gamma^2+4\omega^2)}+\frac{1}{8}\mbox{Re}\left[\frac{1}{\gamma-2\kappa-2i\omega}+\frac{1}{\gamma+2\kappa-2i\omega}\right]\right\},\nn\\
K_3&=&\frac{1}{8\kappa}\left[\frac{4\kappa}{(\gamma^2-4\kappa^2)}+\mbox{Re}\left( \frac{1}{\gamma-2\kappa-2i\omega}-\frac{1}{\gamma+2\kappa-2i\omega}\right)
\right],\nn\\
K_4&=&\frac{\delta}{8\kappa^2}\mbox{Im}\left[\frac{1}{\gamma-2\kappa-2i\omega}+\frac{1}{\gamma+2\kappa-2i\omega}-\frac{2}{\gamma-2i\omega}\right]\nn\\
&=&\frac{\delta\omega}{4\kappa^2}\left[\frac{1}{(\gamma+2\kappa)^2+4\omega^2}+\frac{1}{(\gamma-2\kappa)^2+4\omega^2}-\frac{2}{\gamma^2+4\omega^2}\right].\nn
\eeqna
The remaining coefficients of eq.~(\ref{x2_avg2}) are given by
\beqna
A_{4\omega} &=&\frac{\beta^2 T\gamma}{4\omega^2\kappa^2}\mbox{Re}\left[\frac{1}{\gamma-2\kappa-2i\omega}+\frac{1}{\gamma+2\kappa-2i\omega}-\frac{2}{\gamma-2i\omega}\right],\nn\\
B_{4\omega} &=&\frac{\beta^2 T\gamma}{4\omega\kappa^2}\mbox{Im}\left[\frac{1}{\gamma-2\kappa-2i\omega}+\frac{1}{\gamma+2\kappa-2i\omega}-\frac{2}{\gamma-2i\omega}\right].\nn
\eeqna
Notice that when one gets close to the zone of instability we obtain a far
simpler expression for the average fluctuations. It is given approximately by
\beq
\langle x^2(t)\rangle \approx \frac{2T}{\omega^2}
\left[ \beta^2+\frac{\gamma^2}{4}+\delta^2
\right]\frac{1}{\gamma^2-4\kappa^2}-\frac{4\beta
T\gamma}{\omega^2(\gamma^2-4\kappa^2)}\left[\frac{\delta}{\gamma}\cos(2\omega
t)+\frac{1}{2}\sin(2\omega t)\right]
\label{x2_approx}
\eeq

\section{Linear parametric amplification}
In this section we study the parametric amplification of an added ac signal
near the onset of the first instability zone of Matthieu's equation.
The equation is given by
\begin{equation}
    \ddot x+\omega^2x=-\Omega x-\gamma\dot x+F_p\cos(2\omega
    t)\;x+F_0\cos(\omega t+\phi).
    \label{amp}
\end{equation}
After doing averaging we obtain
\beq
\left(\bea{c}
\dot u \\ \dot v 
\ea
\right)     = -\frac{1}{2\omega}\left[
\begin{array}{cc}
\gamma\omega & \Omega +\frac{F_p}{2}\\
-\Omega +\frac{F_p}{2}&\gamma\omega 
\end{array}
\right]
\left(\bea{c}
u\\v
\ea
\right)+
\frac{F_0}{2\omega}\left(\bea{c}
\sin\phi \\
-\cos\phi
\ea
\right)
    \label{Duffing_amp_av}
\eeq
The fixed points are given by
\[
\left(\bea{c}
u^* \\  v^*
\ea
\right)
=
\frac{F_0}{\gamma^2\omega^2+\Omega^2-F_p^2/4}\left[
\begin{array}{cc}
\gamma\omega & -\left(\Omega +\frac{F_p}{2}\right)\\
\Omega -\frac{F_p}{2}&\gamma\omega 
\ea
\right]
\left(\bea{c}
\sin\phi \\
-\cos\phi
\end{array}
\right)
\]

\beq \left(\bea{c} u^* \\  v^* \ea \right) =
\frac{F_0}{\gamma^2\omega^2+\Omega^2-F_p^2/4}
\left(
\begin{array}{c}
\gamma\omega\sin\phi+(\Omega+F_p/2)\cos\phi\\
(\Omega-F_p/2)\sin\phi-\gamma\omega\cos\phi
\end{array}
\right)
\label{fixed_points}
\eeq
The gain of the amplifier is defined in \cite{almog06} as
\beqna
G(\phi) &=& 20\log\left|\frac{X_{\mbox{pump on}}}{X_{\mbox{pump off} } }\right|
\label{gain}\\
&=&10\log\left\{\frac{\left[(\gamma\omega)^2+\Omega^2+F_p^2/4+F_p\left[\Omega\cos(2\phi)
+\gamma\omega\sin(2\phi)\right]\right](\gamma^2\omega^2+\Omega^2)}{(\gamma^2\omega^2+\Omega^2-F_p^2/4)^2}
\right\}\nn,
\eeqna
where $\left|X\right|=\sqrt{u^{*2}+v^{*2}}$ and the pump off means $F_p=0$.
Rugar and Gr\"utter \cite{rugar91} have studied this amplification process
experimentally and also analytically via a perturbative method described by
Louisell \cite{louis60}. Although, their results agreed well with their
experimental data, we believe that we can increase the applicability of
mechanical parametric amplification of small signals by applying the averaging
method and allowing for detuning. 
We also compare our analytical estimates of gain to the gain obtained from a
full numerical integration of the equations of motion
(\ref{amp}).
The numerical gain is given by the  expression  in Eq.~(\ref{gain}) with the
analytical fixed-point values replaced by the numerical fixed points of the
first-return Poincaré map obtained from the integration of Eq.~(\ref{amp})
after transients died out, i.e. $x(nT)^2+y(nT)^2/\omega^2$, where
$T=\pi/\omega$ and $n>>1$.
\section{Signal to noise ratio}
Following the previous definition of gain, we define a measure of the SNR as
\beq
SNR_0=10 \log \left(\frac{u^{*2}+v^{*2}}{\overline{ \langle x^2 \rangle
}}\right),
\label{snr0}
\eeq
where the fixed points $u^*$ and $v^*$ are given by Eq.~(\ref{fixed_points}) and
the time-averaged thermal fluctuations $\overline{ \langle x^2 \rangle }$ are
given by Eq.~(\ref{x2_avg_dc2}). Near the first instability zone we can write
down this expression, with the help of Eq.~(\ref{x2_approx}), approximately as
\beq
SNR_0=10 \log \left\{
\frac{2F_0^2\omega^2}{T(\gamma^2\omega^2+\Omega^2-F_p^2/4)}
\left[1+\frac{F_p\left[\Omega\cos(2\phi)+\gamma\omega\sin(2\phi)\right]}{(\gamma\omega)^2+\Omega^2+F_p^2/4}\right]
\right\} 
\label{snr0approx}
\eeq
Another measure of the SNR is given by comparing the signal intensity with the noise 
level at the same frequency, that is at $2\omega$. This is given by the expression
\beq
SNR_{2\omega}=10 \log
\left(\frac{u^{*2}+v^{*2}}{\sqrt{|A_{2\omega}|^2+|B_{2\omega}|^2}}\right),
\label{snr2w}
\eeq
where the coefficients $A_{2\omega}$ and $B_{2\omega}$ are defined in Eq.~(\ref{x2_avg2}).
By dimensional analysis one notices that $A_{2\omega}$ and $B_{2\omega}$ have
the same dimensional units as $u^{*2}$ and $v^{*2}$. 
Near the first instability zone we obtain a simple estimate for this SNR
measure, namely
\beq
SNR_{2\omega}=10 \log \left\{
\frac{F_0^2\omega^2}{2|F_p|T(\gamma^2\omega^2+\Omega^2-F_p^2/4)}
\left[\frac{(\gamma\omega)^2+\Omega^2+F_p^2/4+F_p\left[\Omega\cos(2\phi)+\gamma\omega\sin(2\phi)\right]}{
\sqrt{ \Omega^2+\gamma^2\omega^2}}\right]
\right\} 
\label{snr2w_approx}
\eeq
\section{Results and discussion}

\begin{figure}[h]
    \includegraphics[scale=1.0]{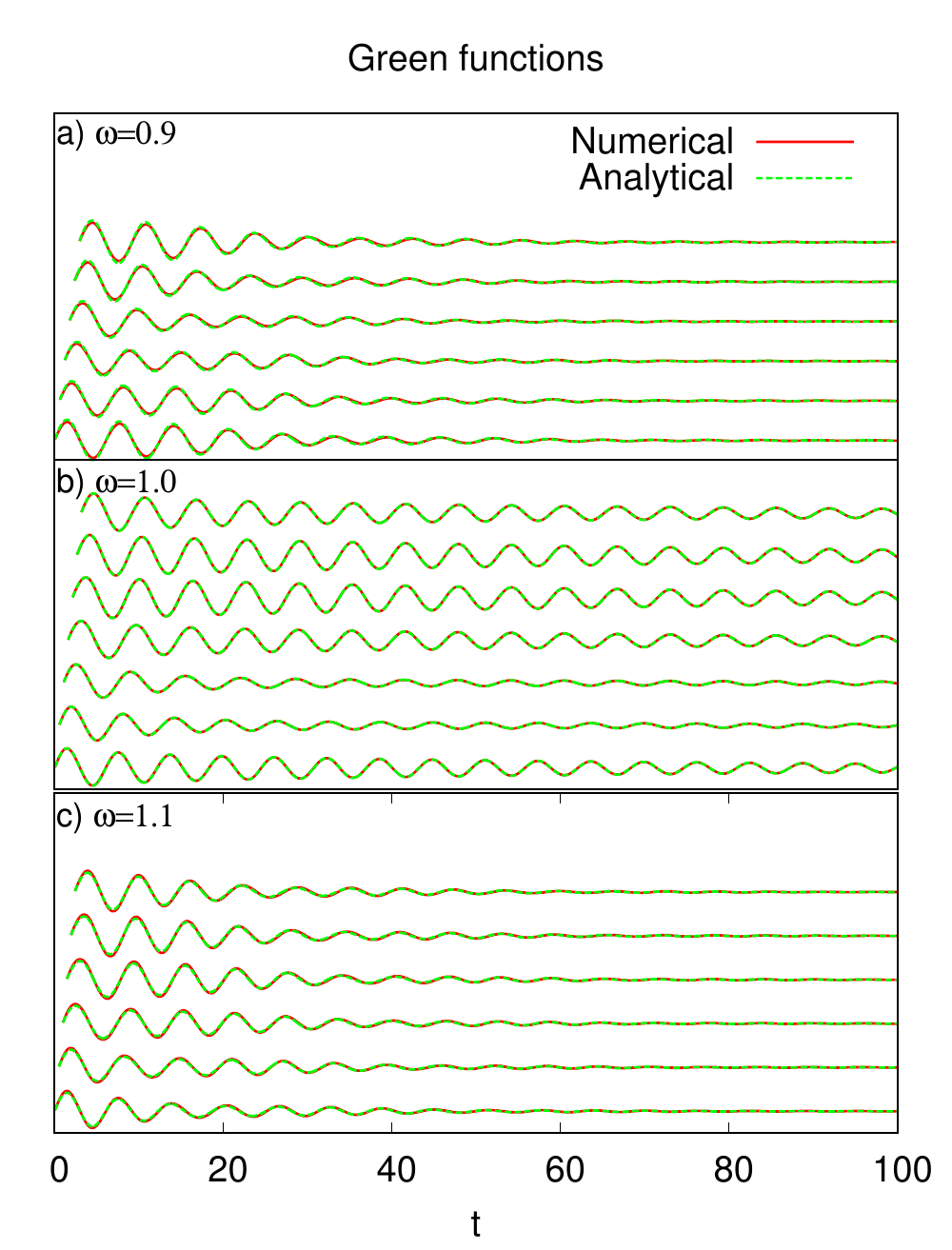}
    \caption{
    (color online) In the frames above we show several Green's functions with
    equally-spaced in time initial conditions in one given period of the
    parametric driving. They are vertically spaced for clarity, since all
    asymptotes are zero. In each frame we have a comparison between numerical
    results given by the numerical integration of Eq.~(\ref{green_eq}) and the
    analytical approximate results given by Eqs.~(\ref{green_avg}) or
    (\ref{green_avg2}). We have a) $\omega=0.9\,\omega_0$, b)
    $\omega=1.0\,\omega_0$, and c) $\omega=1.1\,\omega_0$. The initial values of
    the Green's functions are $G(t,t')=0$ and $\frac{\partial}{\partial
    t}G(t,t')=1.0$ when $t=t'+0^+$. The pump amplitude used was $F_p=0.15$.}
    \label{greenfs}
\end{figure}

In Fig.~(\ref{greenfs}) we plot the Green's functions obtained directly from the
numerical integration of Eq.~(\ref{green_eq}) alongside analytical approximation
results given by Eq.~(\ref{green_avg}), if $|\beta|>|\delta|$, or by
Eq.~(\ref{green_avg2}) if $|\beta|<|\delta|$. We obtain excellent agreement
between the two methods, what implies that our analytical estimates of $\langle
x^2(t)\rangle$ are accurate. The numerical integration was performed using a RK4
algorithm with a time step given by $\pi/(512\omega)$.

\begin{figure}[!h]
    \includegraphics[scale=0.8]{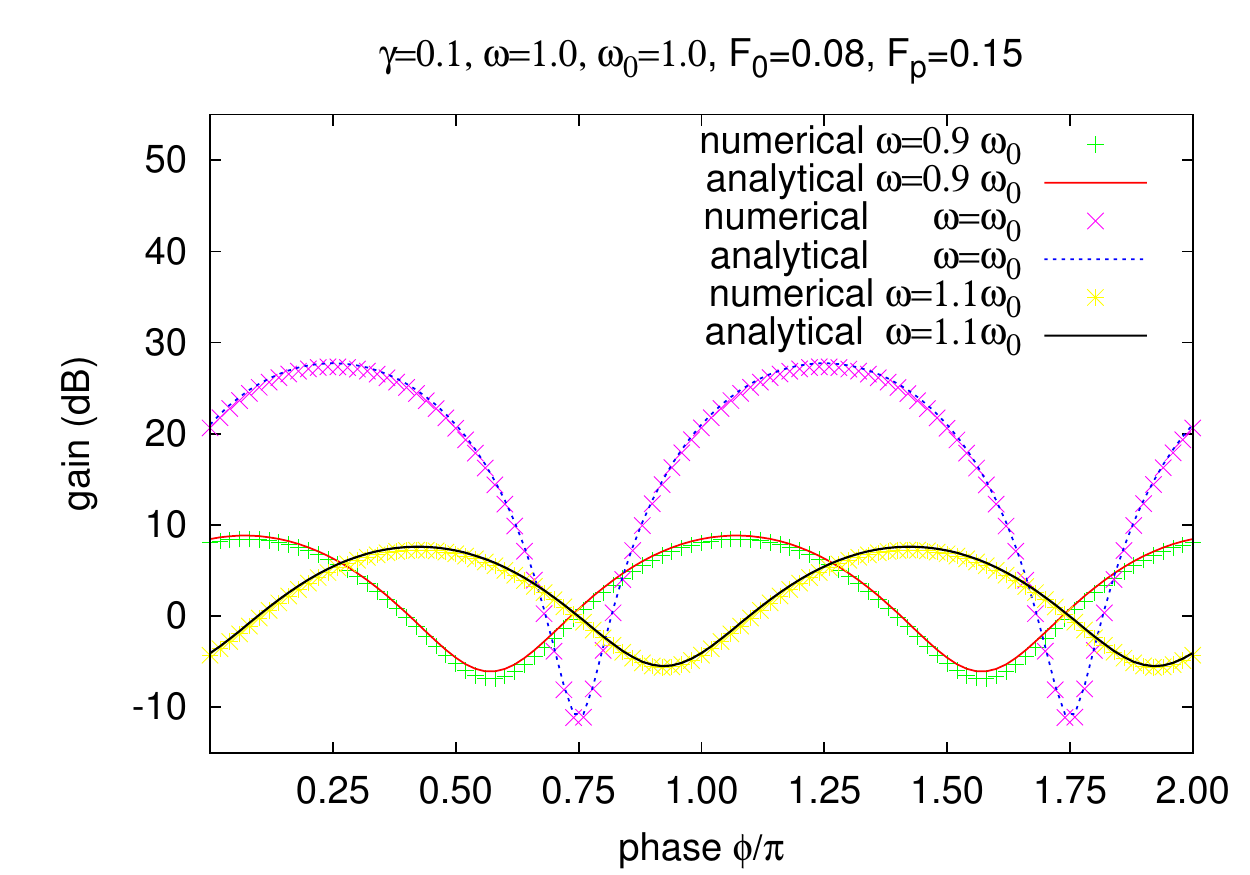}
    \caption{
    Comparison between numerical and analytical estimates of gain as a function
    of phase. The analytical gain is given by Eq.~(\ref{gain}). 
    The numerical values are given by the  expression  in Eq.~(\ref{gain}) with the
    analytical fixed-point values replaced by the numerical fixed points of the
    first-return Poincaré map obtained from the integration of Eq.~(\ref{amp})
    after transients died out and in accordance with the transformation in
    Eq.~(\ref{slow_trans}). Observe that both gain and absorption are reduced
    with detuning.
    \label{fig_gain}
    }
\end{figure}

In Fig.~(\ref{fig_gain}) we obtain excellent fitting between analytical
results for gain obtained by the averaging method in Eq.~(\ref{gain}) and
numerical results given by the fixed point of the first-return Poincaré map of
Eq.~(\ref{amp}) obtained from the integration after transients died out.

\begin{figure}[!h]
    \includegraphics[scale=0.8]{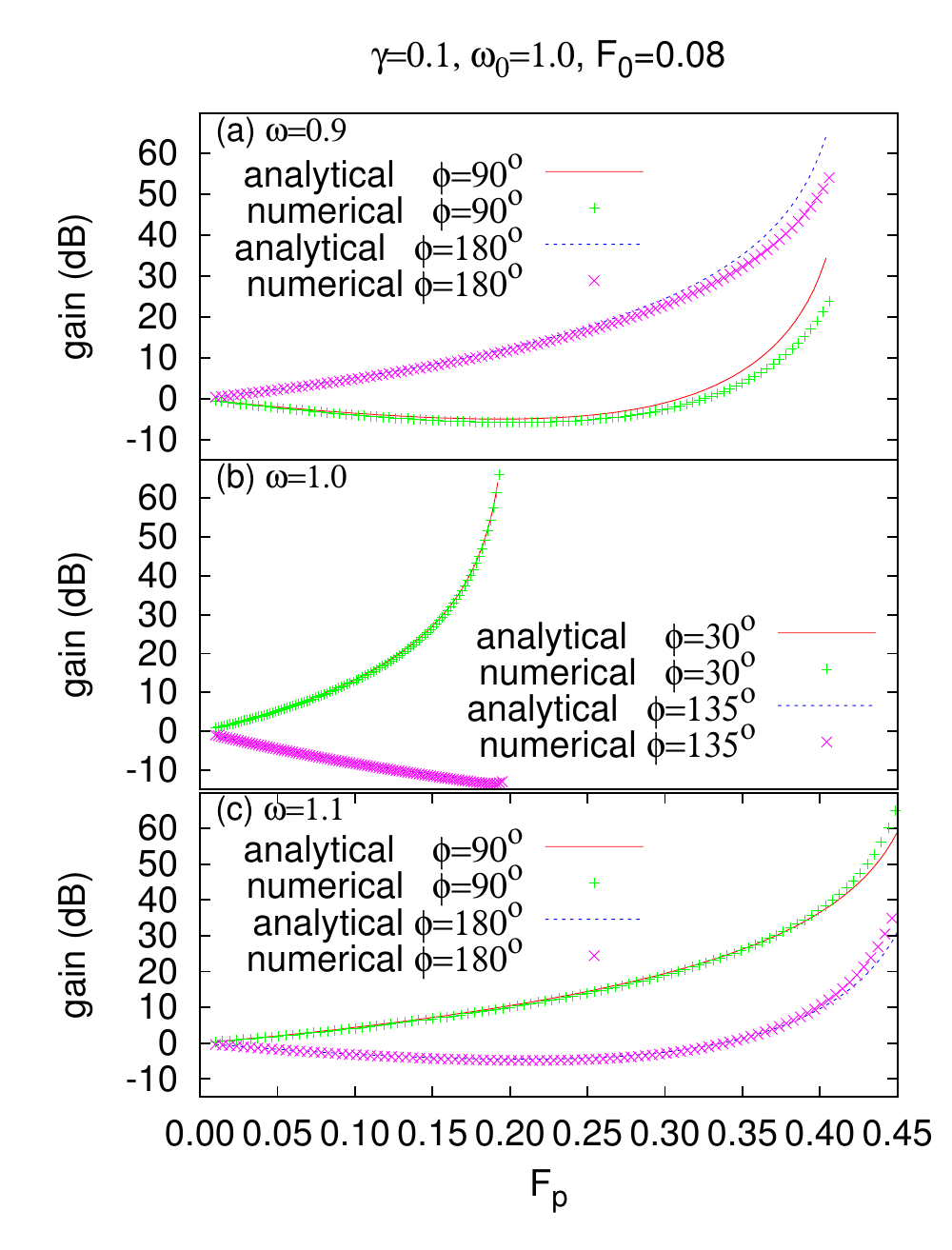} 
    \caption{ Comparison between
    numerical and analytical estimates of gain as a function of pump amplitude
    $F_p$. The analytical gain is given by Eq.~(\ref{gain}). The numerical
    values are given by the  expression  in Eq.~(\ref{gain}) with the analytical
    fixed-point values replaced by the numerical fixed points of the
    first-return Poincaré map obtained from the integration of Eq.~(\ref{amp})
    after transients died out. 
    }
    \label{fig_gain2}
\end{figure}

In Fig.~(\ref{fig_gain2}) a comparison between numerical and analytical
estimates of gain as a function of pump amplitude $F_p$ is shown. Tow
different values of phase are depicted. One observes a very strong dependance on
phase between the pump and the external additive drive. One should obtain a
divergence in gain as the boundary between stable and unstable response is
reached. Again very good estimates are obtained.

\begin{figure}[h]
    \includegraphics[scale=0.7]{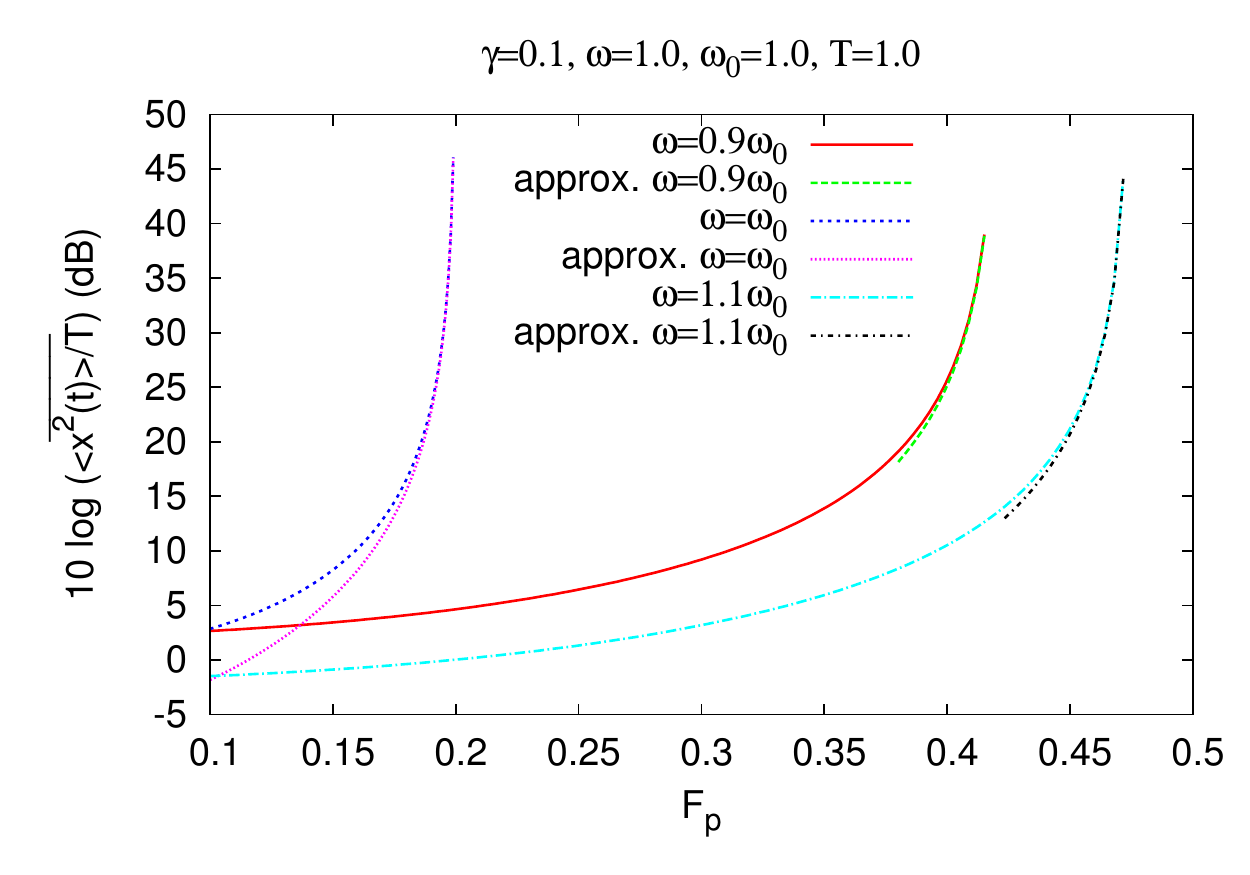}
    \caption{
    (color online) Log plot of the dc component of the mean square displacement,
    $\overline{\langle x^2(t)\rangle}$, as given by
    Eqs.~(\ref{x2_avg_dc2}). In the linear parametric oscillator
    with thermal noise the temperature of the oscillator grows monotonically
    until it diverges at the boundary between stable and unstable zones.
    The approximating curves are given by Eq.~(\ref{x2_approx}).
    }
    \label{tempcont}
\end{figure}

In Fig.~(\ref{tempcont}) we show a logarithmic plot of the dc component of the
mean square displacement over the heat bath temperature. The steep rise of the
curve as the pump amplitude is increased indicates that the noise is also
amplified by the parametric oscillator.

\begin{figure}[!htb]
    \includegraphics[scale=0.8]{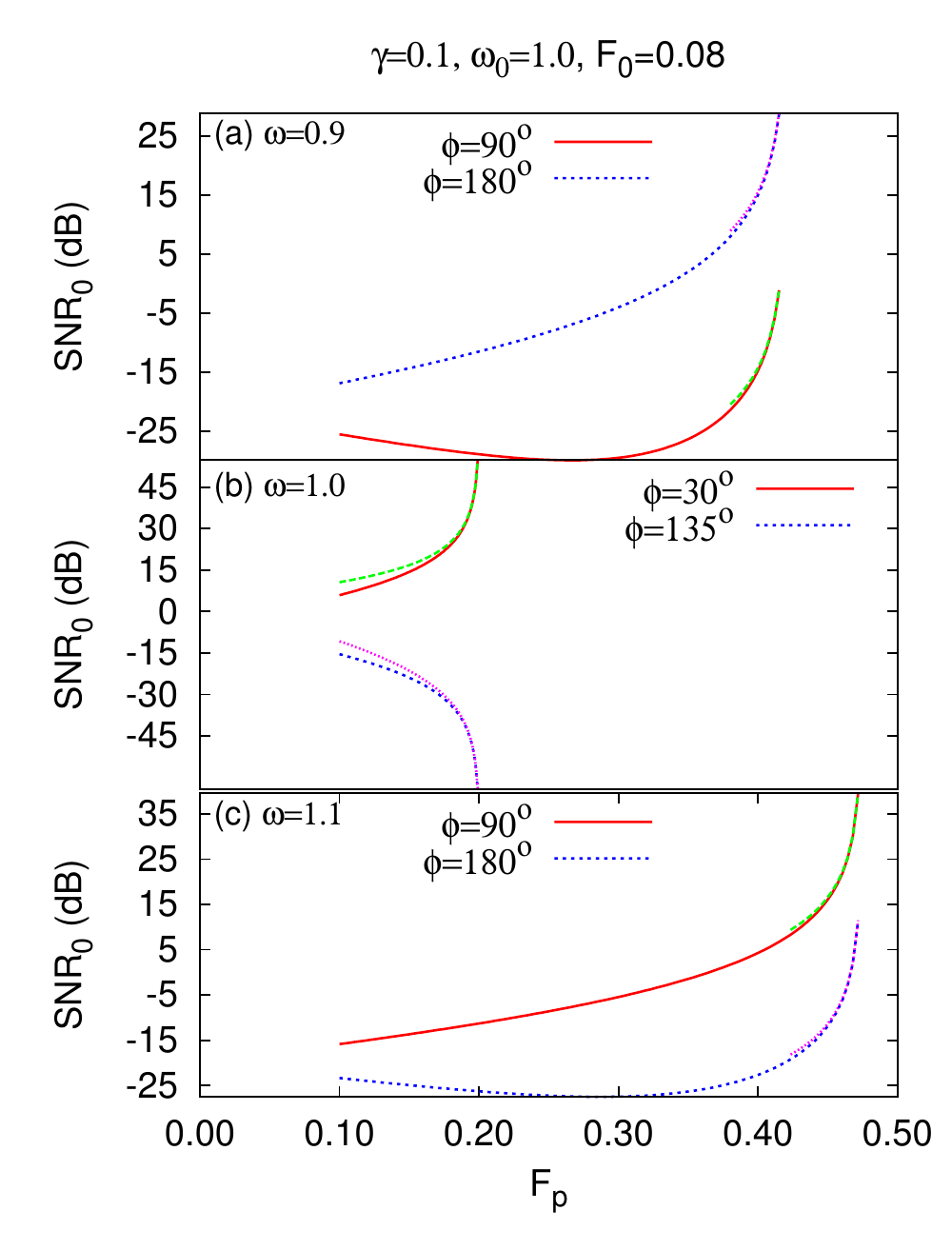} 
    \caption{ 
    Plot of  signal to
    noise ratio SNR$_0$ as defined in
    Eq.~(\ref{snr0}) along with corresponding approximations given by
    Eq.~(\ref{snr0approx}).
    }
    \label{fig_snr}
\end{figure}

In Fig.~(\ref{fig_snr}) we show  logarithmic plots of the amplitude of the
signal (amplitude of response of the parametric oscillator due to the external
ac drive) over the noise (here the dc component of the mean square
displacement). Note that the simple estimates given by Eq.~(\ref{snr0approx})
give very good approximation as one gets close to the first instability zone in
parameter space.

\begin{figure}[!htb]
    \includegraphics[scale=0.8]{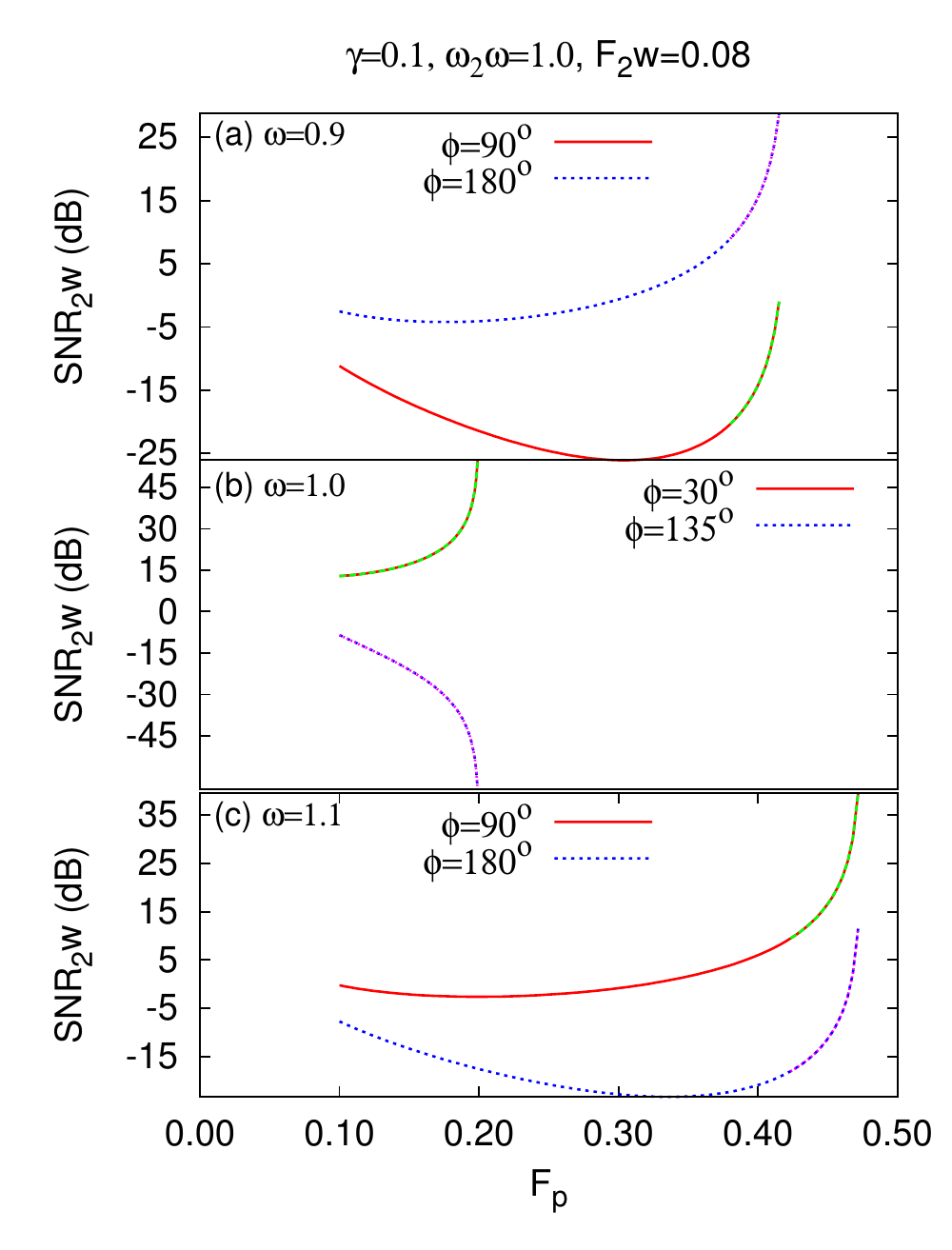} 
    \caption{Plots of the signal to
    noise ratio SNR$_{2\omega}$ as defined in Eq.~(\ref{snr2w}), in which the
    signal is given by $u^{*2}+v^{*2}$ and the noise is given by the amplitude
    of squeezing at $2\omega$ of the mean square displacement $\langle
    x^2(t)\rangle$ as given by $\sqrt{|A_{2\omega}|^2+|B_{2\omega}|^2}$, in
    which $A_{2\omega}$ and $B_{2\omega}$ are defined in Eq.~(\ref{x2_avg2}).
    Very precise approximations are also plotted, given by
    Eq.~(\ref{snr2w_approx}) }
    \label{fig_snr2w}
\end{figure}

In Fig.~(\ref{fig_snr2w}) we show logarithmic plots of the amplitude of the
signal (amplitude of response of the parametric oscillator due to the external
ac drive) over the average fluctuations amplitude that oscillate at frequency
$2\omega$ (here the ac component of the mean square displacement).


\clearpage
\section{Conclusion}
The ability to reduce the influence of thermal noise in parametric amplifiers
(which includes the linear regime of MEMS devices and optomechanical cavities),
can greatly improve the accuracy and precision of measuring small masses and
weak forces.
Here we extended the theoretical work related to the seminal experimental
research  by Rugar and Grütter \cite{rugar91}. For a long time since its
publication there has not appear in the literature a sound explanation based on
stochastic dynamics of the essential features of the classical thermal noise
squeezing phenomena observed there. Recently, though, one of the authors (A.A.B)
has proposed a stochastic dynamics model \cite{batista2011cooling} obtained by
approximating the Green's functions of the parametric oscillator using averaging
techniques to account for the observed experimental effects. Here we extend this
work and give analytical approximate results for SNR in parametric
amplification. Here we propose two different kinds of SNR (SNR$_0$ and
SNR$_{2\omega}$). These estimates of SNR give a measure of the effectiveness of
the parametric amplifiers in the presence additive noise. The analytical results
presented here confirm, that in both measures of SNR, that the parametric
amplifier is indeed a very good amplifier, with sensitive amplification
dependant on frequency, on phase and with low noise.
Further refinement of our results may be achieved by including more details about
the noise model, such as memory effects \cite{grifoni98, batista08}, by taking
into account the coupling to a heat bath that could be made out of photons,
as in radiation-pressure cooling, or via coupling to phonons. 
Further improvements of the accuracy of the predictions should be obtained by
taking nonlinear terms into account, especially when one gets close, in
parameter space, to the first zone of instability.
It is noteworthy to observe that our method may be applied to the linear
response of nonlinear oscillators in the presence of both an ac drive and thermal noise.

Finally, we note that this model can also be applied to the dynamics of ions in
quadrupole ion guides or traps \cite{paul90} or traps for neutral particles with
magnetic dipole moments \cite{lovelace1985}. The presence of noise would
indicate that the vacuum is not complete. One would obtain an estimate of the
limits of mass spectroscopy in quadrupole ion guides, where the spectral limit is
bounded by the ion guide length and the amount of noise present in the system.

%

\clearpage

\end{document}